\documentclass[aps,amsmath,prb,twocolumn,a4paper,showpacs,floatfix]{revtex4}
\usepackage{graphicx}
\newcommand{\comment}[1]{}
\newcommand{\beq}{\begin{equation}}
\newcommand{\eneq}{\end{equation}}
\newcommand{\bea}{\begin{eqnarray}}
\newcommand{\enea}{\end{eqnarray}}
\newcommand{\bc}{\begin{center}}
\newcommand{\ec}{\end{center}}
\newcommand{\vt}{ \varphi}
\newcommand{\ct}{ \cos  \Theta}\newcommand{\st}{ \sin  \Theta}

\newcommand{\met}{\frac{1}{2}}

\newcommand{\df}{\dot {\varphi}}
\newcommand{\dP}{\dot {\Phi}}
\newcommand{\dTh}{\dot {\Theta}}

\newcommand{\cmt}{ \cos \frac{ \vartheta}{2}}
\newcommand{\smt}{ \sin \frac{ \vartheta}{2}}

\newcommand{\smqt}{ \sin ^2\frac{ \vartheta}{2}}

\newcommand{\wo}{\omega_o}

\newcommand{\h}{\hbar}
\newcommand{\bd}{\begin{displaymath}}
\newcommand{\ed}{\end{displaymath}}

\newcommand{\ang}{ i\omega _o t }
\newcommand{\bwt}{\begin{widetext}}
\newcommand{\ewt}{\end{widetext}}
\begin{document}

\title{Quantum Rings with Rashba spin orbit coupling: a path integral
approach}

\author{P. Lucignano$^{1,2}$}\author{D. Giuliano$^{1,3}$} 
\author{A. Tagliacozzo$^{1,2}$}

\affiliation{$^1$ Dipartimento di Scienze Fisiche Universit\`a degli
             studi di Napoli "Federico II", Napoli, Italy}
\affiliation{$^2$ Coherentia-INFM,
             Monte S.Angelo - via Cintia, I-80126 Napoli, Italy}
\affiliation{$^3$ Dipartimento di Fisica, Universit\`a della Calabria and
             I.N.F.N., Gruppo collegato di Cosenza, Arcavacata di Rende
             I-87036, Cosenza, Italy}

\date{\today}
\begin{abstract}
We employ a path integral real time approach to compute the DC
conductance and spin polarization for electrons transported across a
ballistic Quantum Ring with Rashba spin-orbit interaction.  We use a
piecewise semiclassical approximation for the particle orbital motion
and solve the spin dynamics exactly, by accounting for both Zeeman
coupling and spin-orbit interaction at the same time. Within our
approach, we are able to study how the interplay between Berry phase,
Ahronov Casher phase, Zeeman interaction and weak localization
corrections influences the quantum interference in the conductance
within a wide range of externally applied fields. Our results are
helpful in interpreting recent measurements on interferometric rings.
\end{abstract}

\maketitle

\section{Introduction}

In classical physics, a charged particle moving in an external
magnetic field $B$ feels a (Lorentz) force only in the regions in
which $B$ is different from zero. Instead, in 1959 Ahronov and Bohm
(AB)\cite{aharonovbohm} showed that the wave packet representing the
state of a quantum particle can be influenced by an external vector
potential $\vec{A}$, even if the corresponding magnetic field is zero,
provided that the particle is moving in a space with a nontrivial
topology (holes). To be more specific, the wavefunction of the charged
particle may acquire a nonzero phase, when undergoing a closed path in
a space threaded by an external magnetic flux.

Nowadays, mesoscopic quantum rings (QR's) allow to have direct access to
the phase of the electron wavefunction, since their size is smaller
than the distance over which the phase randomizes, as a consequence of
scattering against impurities and interactions. The total rate of
particles coming from the two arms of the ring and interfering at the
exit contact can be directly probed by measuring the QR conductance.
Indeed, interference effects have been observed in metal QRs many years
ago\cite{webb}.  In addition, since electrons are spinful particles,
the spin part of the wavefunction is influenced by the magnetic field
via the Zeeman term in the Hamiltonian.  Moreover, a more subtle
effect arises when there is a magnetic field non-orthogonal to the
plane of the orbiting particle because, as a consequence of the
orbital motion, its spin dynamics is instantaneously governed by a
time-dependent Hamiltonian \cite{anandans}.  This time dependence ends
up in an extra phase acquired by the particle wavefunction which is
named after Berry
\cite{berry,anandan}, who put in foreground its topological properties
when the orbits are closed.

Recently, semiconductor technology allows to grow samples (for
instance made by InAs or InGaAs) with sizeable spin orbit interaction
(SOI). Rashba\cite{rashba} first pointed out that the SOI strength can be
controlled by means of voltage gates.  This feature has been recently
found experimentally \cite{meijer,miller}. Because of the SOI, an electric 
field $E$ orthogonal to the orbit provides a momentum dependent effective 
magnetic field felt by the electron spin in orbital motion.  Such a
field can be equally well described by a vector potential that adds an
extra phase to the wavefunction\cite{aronov}. This phase was spelled
out by Ahronov and Casher \cite{ac}, as a ``dual'' AB effect, with
charge and spin interchanged (together with $B$ and $E$).

During the last years, the effects of SOI on the AB oscillations have
been observed in semiconductor based QR by several groups
\cite{yau,nitta,morpurgo,kato}.  It has been recognized that, in the
presence of both $B$ and $E$ orthogonal to the orbit plane, the
effective, momentum dependent  total $B$ field is tilted w.r. to the vertical
direction. The resulting Berry phase influences the interference
pattern. Such a result is of the utmost interest because the Rashba-SOI
(RSOI) turns out to be a tool to tune the Berry phase and, ultimately,
the conductance, as well as the spin polarization of the outgoing
electrons.  In Ref. \cite{molenkamp}, it is clearly shown that the AB
interference fringes can be modified by tuning an external
electrostatic potential.

In a recent publication\cite{noiletter} we have included all these
phenomena affecting the interference of an electron ballistically
transported in a ring, also accounting for some dephasing at the
contacts. We have also shown that the AB peak in the Fourier transform
of the magnetoconductance displays satellite peaks due to the RSOI. It
has been suggested that satellite peaks observed in recent
experiments\cite{shayegan} have the same origin. In the present work
we review the theory by presenting the full calculation. In addition,
we include the semiclassical paths leading to weak localization
corrections\cite{nonloso}, and we discuss the rotation of the spin
polarization, during the transport across the ring.

During the last years, several theoretical techniques have been
employed to study quantum transport in a QR. In Ref.s \cite{loss}, an
imaginary time path integral approach\cite{feynman} is developed to
study the conductance of a strictly one dimensional (1D) QR and its
conductance fluctuations in the diffusive limit. In
Ref. \cite{tserkov} a real time path integral approach is applied in
the limit of negligible Zeeman splitting.  Several papers have
discussed the conductance properties and the spin selective transport
of QR's in the strictly 1D ballistic limit, by means of a spin
dependent scattering matrix approach
\cite{molnar,frustaglia,shen,dario,citro}. In the absence
of the magnetic flux, the conductance shows quasi-periodic
oscillations in the SOI stength, which can be modified by switching
the magnetic field on.  Numerical calculations \cite{souma,lozano,wu}
have shown that in the 2D case there are only quantitative
modifications of the 1D results that do not qualitatively affect the
physics.

In this paper we extend the real time path integral approach
previously developed in Ref.\cite{noiletter} to study the conductance
and the spin transport properties of a ballistic quantum ring in the
presence of both RSOI and of an external magnetic flux orthogonal to
the ring plane. We use a ``piecewise'' saddle point approximation for
the orbital motion, keeping the full quantum dynamics of the
spin. This approach allows us to take into account, in a
nonperturbative way, both the RSOI and AB phase and to include also
the Zeeman spin splitting.

Our numerical approach evaluates all paths contributing to the quantum
propagator. The scattering at the leads can be forward or backward,
according to the probability amplitudes given by the S-matrix. Weak
localization corrections can be easily extracted from our result. We
also allow for some diffusiveness at the contacts by adding a random
phase factor in the motion.

The DC conductance is derived from the Landauer formula
\cite{landauer} $ {\cal{G}} ={e^2}/{\h} \sum_{\sigma\sigma'}
\left|A(\sigma; \sigma'|E)\right|^2$, where $A(\sigma ;\sigma'|E)$ is
the probability amplitude for an electron entering the ring with
energy $E$ and spin polarization $\sigma '$ to exit with spin
polarization $\sigma $. We also report the change in the spin
polarization the electron transported across the ring.

The structure of the paper is as follows:
\begin{itemize}
\item
In Section II we introduce the Feynman propagator for a spinful
electron injected at the Fermi energy in the ring.
\item 
In Section III we discuss the topology of the allowed paths and the
scattering of the electron at the leads.
\item
In Section IV we represent our path integral in the coherent spin
basis\cite{haldane,plet} and derive the saddle point equations of motion,
whose classical counterpart is described in detail in Appendix A. This
allows us to justify the choice of a piecewise semiclassical
approximation for the orbital motion of the electron in the ring.
\item
In Section V we present the details of the calculation by rewriting
the path integral as a collection of single arm propagators. These are
the building blocks to be calculated in the next section.
\item
In Section VI we analyze how the orbital motion affects the full
quantum dynamics of the electron spin for each arm of the ring and
chirality. The spin propagator is derived in Appendix B, in the basis
corresponding to the rotating reference frame in the spin space.
\item
In Section VII we discuss the dependence of the conductance on the external
fields and on the overall transmission across the ring.
\item
In Section VIII we focus on the spin polarization of the outcoming
electron.
\item
Section IX includes a short summary and our conclusions.
\end{itemize} 
\section{The transmission  amplitude}

Our model Hamiltonian will be the two-dimensional Hamiltonian for a
particle with spin-1/2 $\vec{S}$, in an orthogonal magnetic field,
with spin-orbit coupling to an orthogonal electric field (Rashba
coupling).  It is given by
\bea
H[ \vec{p} , \vec{r} , \vec{S} ]  =
\frac{1}{2m}\left(\vec p+\frac{e}{c}\vec A_0\right)^2
-  \omega_{c}\: S_{z}+{H}_{so}
 \label{hamilt} \\
 H_{so}= \frac{2\alpha}{\hbar ^2}\left( {\hat z} {\times} 
\left({\vec p+\frac{e}{c} \vec A_0}\right) \right){\cdot}
{\vec{S}}  \:\:\:\: , \nonumber
\enea
where $\alpha$ is the spin orbit coupling constant, in units $ eV
\:$\AA, $\vec{S} = \hbar \vec{\sigma } /2 $
(${\sigma_x,\sigma_y,\sigma_z}$ are the Pauli matrices),
$\vec{A}_0(\vec{r}) = \frac{B}{2} ( - y , x , 0 ) $ is the vector
potential generating the uniform field $B$, normal to the ring
surface, taken in the symmetric gauge, $\omega_c = ge B / 2mc$ is the
cyclotron frequency. In real nanostructures based on InAs or InGaAs
the $g$ factor can strongly deviate from the value of two.  However
the result we present here, are fully general as they depend on the
ratio $\alpha/\hbar \omega_c R$ which can be tuned by acting on
$\alpha$.  Since we will assume only a single channel to be avaliable
for electron propagation across the ring, we will picture the single
channel ring as a $1-d$ circle of radius $R$, connected to two
leads. Accordingly, the position of the particle within the ring is
parametrized by the angle $\vt$ and the vector potential has just the
azimutal component $A_\vt = \phi / 2\pi R $, where $\phi$ is the
magnetic flux threading the ring.

In order to study the conduction properties of the ring, one needs the
propagation amplitude for an electron entering the ring with spin
polarization $\mu_0$ to exit with spin polarization $\mu_f$, at energy
$E_0$. This is given by

\beq
A(\mu _f ; \mu _0 | E_0 )  =
\int_{0}^{\infty} \: \frac{d t_f}{\tau _0}  \:  e^{i \frac{E_{0}
t_f}{\hbar}}  \:
\langle {\vec r}_{f} , \mu_{f} , t_{f}|{\vec r}_{0} , \mu_{0 } , t_{0} \rangle
 \:\:\:\: ,
\label{ampclas1}
\eneq

where $\langle {\vec r}_{f} , \mu_{f} , t_{f}|{\vec r}_{0} , 
\mu_{0 } , t_{0} \rangle$ is the  amplitude for  a particle entering the
ring  at the point $\vec{r}_0$ and  at the time $t_0$ with spin 
polarization $\mu_0$ to exit at the point $\vec{r}_f$ at the
time $t_f$ with spin polarization $\mu_f$. In our tensor product notation,
we define $|{\vec r} , \mu \rangle=|{\vec r}\rangle\:\otimes\:|\mu\rangle\ $.
 $\tau _0 =mR^2 /(2\hbar) $ is the
time scale for the quantum motion.

In order to compute $\langle {\vec r}_{f} , \mu_{f} , t_{f}|{\vec
r}_{0} , \mu_{0 } , t_{0} \rangle$ , we resort to a path integral
representation for the orbital part of the amplitude. Since we
parametrize the orbital motion of the particle in terms of the angle
$\varphi$, we provide the pertinent Lagrangian, ${\mathcal{L}}_{orb
}$, as a function of $\varphi , \dot{\varphi}$.  It is given by 
\beq
{\mathcal{L}}_{orb}[\varphi(t), \dot \varphi(t),\vec \sigma]=
\frac{m}{2}R^2\dot\varphi^2(t)-\frac{\phi}{\phi_{0}} \hbar
\dot\varphi(t) + \frac{\alpha^2\:m}{2 \hbar^2}+ \frac{\hbar^2}{8 m
R^2} \:\:\:\: .
\label{lag}
\eneq 
The last two contributions to Eq.(\ref{lag}) are a constant, coming from
the spin-orbit term, and the Arthurs\cite{morette} term, which is
required when a path integration is performed in cylindrical
coordinates. Since both contributions are constant, they can be lumped
into the incoming energy $E_0$ and therefore they will be omitted
henceforth.
By taking into account the spin degree of freedom, as well, we represent
the propagation amplitude as
\bwt  
\beq 
\langle {\vec r}_{f}
, \mu_{f} , t_{f}|{\vec r}_{0} , \mu_{0 } , t_0 \rangle = \langle
{\vec r}_{f} , \mu_{f} | e^{ - i \int_{t_0}^{t_f } \: d t \: H } | {\vec
r}_{0} , \mu_{0 } , t_0 \rangle =\int_{\varphi ( t_0 ) =
\varphi_0}^{\varphi ( t_f ) = \varphi_f} \!\!\!\!{\cal D} \varphi\: e
^{- i \int_{0}^{t_f} d t \; \left[ \tau_0 \dot{\varphi} ^2 - q
\dot{\varphi}\right ]} \: \langle \mu _f | \hat{U}_{spin} ( t_f, t_0 ) |
\mu _0 \rangle \:\:\:\: ,
\label{parte}
\eneq  
\ewt
where $q=\phi / \phi_o $,  $\phi_o $ being the
flux quantum $hc/e$. 
\beq
\hat{U}_{spin} ( t_f, t_0 ) = {\bf T} \exp \left [ -\frac{i}{\hbar} \:
\int _{t_0}^{t_f} dt' \: \hat{H} _{spin} (t') \right ] \:\:\: .
\eneq
is the full  spin  propagator and the spin Hamiltonian  
$ \hat{H} _{spin} (t)$ is given by 
\beq
\frac{1}{\hbar } \hat{H}_{spin} ( t ) =
 \left[ \begin{array}{cc}  \frac{  \omega _c}{ 2}  & 
 \gamma  \dot\varphi  e^{ - i \varphi ( t ) } \\
\gamma  \dot\varphi e^{  i \varphi (t)  } &
-  \frac{ \omega _c}{ 2}  \end{array} \right]
\:\:\:\: , 
\label{three}
\eneq

with $\gamma = 2\alpha \tau _0 /(\hbar R ) $\cite{nota}.
In Sections IV and V we show that the amplitude of Eq.(\ref{parte}) can
be approximated by choosing a piecewise semiclassical orbital motion
for the particle in each arm of the ring, while keeping the full
quantum dynamics of the spin.
In particular, we will see that, within the physically relevant range
of parameters, the orbital motion can be approximated as a uniform
rotation (with constant angular velocity), which makes the spin
dynamics to be the one of a spin-1/2 in an effective, rotating,
external magnetic field.  Yet, in order to explain how we deal with
quantum backscattering at the contacts between ring and arms and
corresponding dephasing effects, we will introduce our formalism in
the next section, by discussing a simplified version of our problem: a
spinless electron propagating across a mesoscopic ring.
\section{Feyman's paths for a Spinless particle transmitted across  a  ring}
\begin{figure}[!htp]
    \centering \includegraphics[width=\columnwidth]{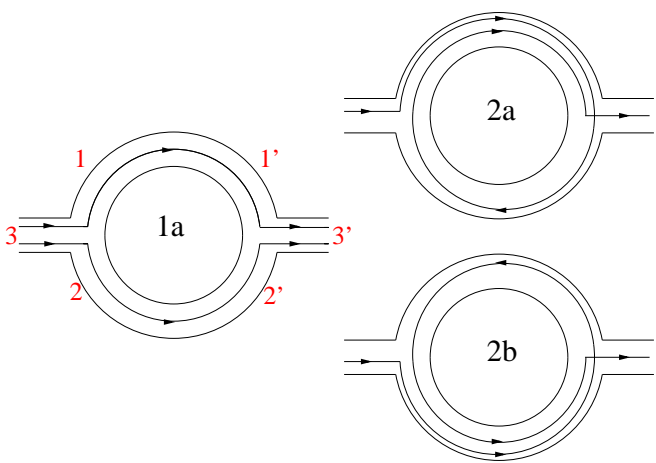}
\caption{(color online) First and second order paths included in the
calculation of the transmission amplitude across the ring, 
from left to right, including
forward scattering only. Numbers $1,2,3(1',2',3')$ in Fig. 1a refer to
the labeling of the terminals in Eq.(\ref{matr}.)}
\label{noWLpaths}
\end{figure}

In this section we introduce our formalism by computing the 
transmission amplitude for  a spinless electron of  mass $m$ and charge  $-e$, 
traveling   across the ring   in  an 
orthogonal magnetic field. 
For a realistic device, at each lead one has to take into account three 
possible scattering processes, consistently with the conservation of the
total current. This is  described in terms of a unitary $S-$matrix
that, in the  symmetric case in which  the two arms are symmetric, is
given by
\beq
\mathcal S=\left(
\begin{array} {c c c}
         -\met (1+\bar r) & \met (1-\bar r)          & \sqrt{\met(1-\bar r^2)}\\
          \met (1-\bar r) &-\met (1+\bar r)          & \sqrt{\met(1-\bar r^2)}\\
 \sqrt{\met(1-\bar r^2)}  & \sqrt{\met(1-\bar r^2)}  &  \bar r
\end{array}\right)
\label{matr}
\:\:\:\: .
\eneq
The numerical labeling of the S-matrix elements referring to the three
terminals of each contact fork, are explained in Fig.(\ref{noWLpaths},
1a).  Assuming, for simplicity, that the scattering matrix is the same
for both leads, Eq.(\ref{matr}) will hold both at the left-hand lead,
and at the right-hand lead of the ring.

In particular, $\mathcal S_{3,3}=\bar r$ is the reflection amplitude
for a wave coming from the left lead, $\mathcal S_{1(2),1(2)}=-\met
(1+\bar r)$ the reflection amplitude for a wave incoming from the
upper/lower arm, $\mathcal S_{1(2),2(1)}= \met (1-\bar r)$ is the
transmission amplitude from the upper (lower) to the lower (upper) arm
and $\mathcal S_{1(2),3}=\mathcal S_{3,1(2)}=\sqrt{\met(1-\bar r^2)}$
is the transmission amplitude from the upper/lower arm to outside of
the ring.

In Fig.(\ref{noWLpaths}) we show the simplest paths of the electrons
in the ring including forward scattering at the contacts, only.
The contribution to the total amplitude coming from such paths, in
which the electron enters the ring at an angle $\varphi_0$ at time
$t_0$ and exits at $\varphi_f$ at time $t_f$, is given by
%
\bea 
{\cal{A}}(\vt_{f}, t_f; \vt _0, t_0  ) =  \sum _{n=-\infty}^{+\infty} 
\int _{\vt (t_0)=
 \vt _i}^{\vt (t_f)= \vt _f +2\pi n }\!\!\!\!\!\!\! {\cal{D}}\vt (\tau ) \: 
e^{-i{\cal{S}}[\vt[t)]/\hbar }\nonumber\\
  =\sum _{n=-\infty}^{+\infty} 
e^{-iq \:(\vt_f-\vt_i +2\pi n ) }
\int _{\vt (t_0 )=
 \vt _i}^{\vt (t_f)= \vt _f +2\pi n }\!\!\!\!\!\!\! 
\!\!\!\!\!\!\! 
\!\!\!\!\!\!\! 
\!\!\!\!\!\!\! 
{\cal{D}}\vt (t ) \: 
e^{-i\frac{mR^2}{2\hbar }  \int _{t_0}^{t_f} dt \: \dot \vt ^2 (t )}  
\:,   
\label{spinless}
\enea
where we have summed over paths in which the electron winds $n+1/2$ times in 
the ring, before exiting it. 
Positive (negative) $n$ values imply clockwise (counterclockwise)
propagation along the ring.  

This propagator can be evaluated exactly
\cite{moran}.  However we report here just the saddle point
evaluation, for comparison with the spinful case.  Minimizing the 
action gives the classical equation of motion (together with the pertinent
boundary conditions for a path winding  $n+1/2$ times): 

\beq 
\ddot \vt (t ) 
= 0 \:\:\:
; \vt (t_i ) = \vt _i \: , \vt (t_f ) = \vt_f + 2\pi n 
\:\:\:\: . 
\eneq 
Let us assume that the particle is injected in the ring at $\vt_0 =0 $
and comes out at $\vt_f = \pi $ in a time $T = t_f$.
Eq.(\ref{spinless}) gives:
\bea {\cal{A}}(\pi ,0, t_f ) =
e^{i\pi q } \: \sqrt{\frac{\tau_0}{\pi i t _f}} \: {\sum
_{n=-\infty}^{+\infty }}' e^{-i \pi ^2 (2|n|-1) ^2 \tau_0  / t_f -i
2\pi n q } \nonumber\\
\:\:\:\: , 
\label{less} 
\enea
where the prime in the sum takes into account the fact that one does not
sum over  $n=0$, and the square root at the prefactor accounts for the
gaussian fluctuations. Of course, this propagator is periodic in $q$ of
period $q=1$ up to a minus sign. 

\begin{figure}[!htp]
    \centering
    \includegraphics[width=\columnwidth]{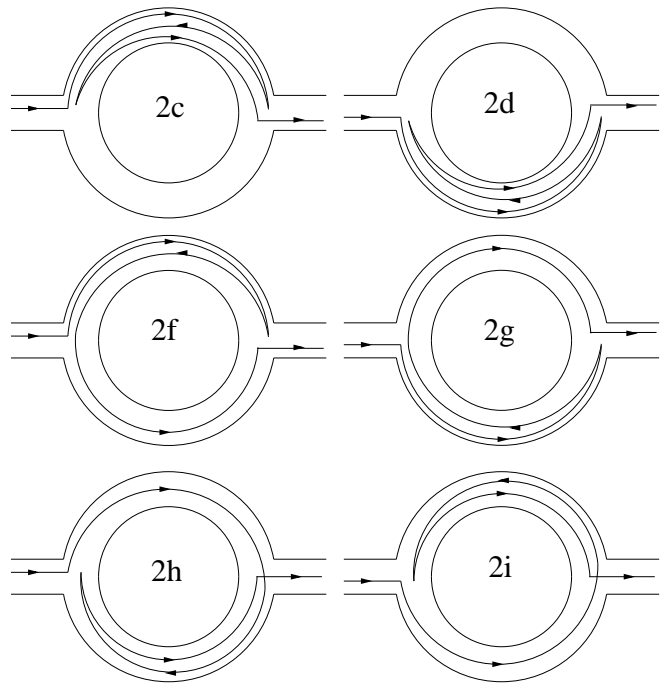}
\caption{Second order paths of the transmission amplitude from left to
  right including backscattering at the leads. Paths $(2f)$ and
  $(2h)$, as well as $(2g)$ and $(2i)$ contribute to the weak
  localization corrections.}
\label{WLpaths}
\end{figure}

More involuted paths arise if one takes into account backscattering
processes in which the electron can get backscattered within the same
ring's arm from which it is coming. For instance, the paths $(2f)$ and
$(2h)$, as well as $(2g)$ and $(2i)$ in Fig.(\ref{WLpaths}), include
looping in opposite directions around the ring's hole. Interference
between clockwise and counterclockwise windings leads to weak
localization corrections. We denote these corresponding paths
(including also $(2c)$ and $(2d)$ ) as ``reversed paths''$(RP)$. In our
approach, all order paths are numerically generated up to the
convergency and the $S-$matrix of Eq.(\ref{matr} ) is implemented in
the numerical algorithm.
In conclusion, we have established that the paths contributing to the
transmission amplitude can be built up by adding four types of
elementary paths: $u_\rightarrow$ forward orbiting in the upper arm of
the ring ($\vt \in ( 0,\pi )$), $ u_\leftarrow$ backward orbiting in
the upper arm of the ring ($\vt \in ( \pi,0 )$), $d_\rightarrow$
forward orbiting in the lower arm of the ring ($\vt \in ( 2 \pi,\pi )$), $
d_\leftarrow$ backward orbiting in the lower arm of the ring ($\vt \in
( \pi, 2 \pi )$). In Section V, we will generalize such an approach, while
in the next Section we discuss the Feynman path-integral
representation of the propagation amplitude for a spinful electron
propagating along one of these elementary paths.
\section{Quantum amplitude and semiclassical  orbital  motion 
for a spinful electron in the ring}
In this section, we construct the Feynman representation in the basis
of the coeherent spin states, for the propagation amplitude of an
electron moving along one of the arms of the ring with a given
chirality.  To discuss the saddle-point approximation we need the
equations of motion coming from the condition that the action is
stationary.  The coherent spin state basis provides a straightforward
route to perform a semiclassical approximation involving both orbital,
and spin degrees of freedom, at the same time. In particular, we will
show under which conditions the classical orbital motion $\dot{\vt}=
cnst $, can be retained, as in the spinless case.  Accordingly, the
spin dynamics will be that of a ``quantum magnetic moment'' in a
time-dependent external magnetic field.

Let $\Omega$ denote the orientation of the spin $\vec S$ and $|\Omega
\rangle$ be the coherent state such that:
\beq
\langle \Omega | \hat{\vec{S}}|  \Omega \rangle  \equiv \vec{S}[\Omega ] = 
\hbar \: S \:  \left[ \begin{array}{c} \sin  \Theta  \cos  \Phi  \\
\sin  \Theta  \sin  \Phi  \\ \cos  \Theta  \end{array} \right]
\:\:\:\: , 
\label{ringo2.2.2}
\eneq
\noindent
for the three components of the spin vector, respectively. 
The full propagator in the coherent spin state representation is given
by:
\bwt
\bea
\langle \varphi_f , \Omega _f , t_f \: |  \varphi_0 , \Omega _0 , 0
\rangle =\int_{\varphi ( 0 ) = \varphi_0}^{\varphi ( t_f ) = \varphi_f} 
\!\!\!\!{\cal D} \varphi\:   e ^{- i \int_{0}^{t_f} d t \;
 \left[  \tau_0  \dot{\varphi} ^2 -  q \dot{\varphi}\right ]}
\int_{\Theta ( 0 ) = \Theta_0}^{\Theta ( t_f )
 = \Theta_f} \!\!\!\!
{\cal D} \Theta \int_{\Phi ( 0 ) = \Phi_0}^{\Phi ( t_f ) = \Phi_f} 
\!\!\!\!{\cal D} \Phi\: 
\: e^{-\frac{i}{\hbar} 
{\cal{S}}_{spin}\left [ \Theta , \dot{\Theta} ; \Phi , 
\dot{\Phi} | \vt , \dot{\vt} \right ] }
\:\:\:\: , 
\label{propo}
\enea
where:
\bea
{\cal{S}}_{spin} \left [ \Theta , \dot{\Theta} ; \Phi , 
\dot{\Phi} | \vt , \dot{\vt} \right ]  /\hbar  = \int_{0}^{t_f} d t \;
 \left \{ \frac{ (1-\cos\Theta )}{2} \dot \Phi + {\cal {L}}_{spin} 
 \left [ \Theta , \dot{\Theta} ; \Phi , 
\dot{\Phi}| \varphi , \dot{\varphi}   \right ] \right \} 
\:\:\:\: . 
\label{prospin}
\enea

with:
\bea
  {\cal L}_{spin}  \left [ \Theta , \dot{\Theta} ; \Phi , 
\dot{\Phi} | \varphi , \dot{\varphi} \right  ] =
  \frac{ \omega_c}{2} \cos \Theta - \gamma
\dot{\varphi} \sin \Theta \cos ( \varphi - \Phi  ) 
\:\:\:\: . 
\label{venti}
\enea

\ewt
The Lagrangian of Eq.(\ref{venti}) corresponds to the coherent
spin-state representation of the classical Lagrangian derived from
Eq.(\ref{hamilt})
\bea
 {\cal L} [ \varphi , \dot{\varphi} , \vec{S}  ] =
 \nonumber \\
 \frac{m}{2}   \dot\vec{r} ^2  - 
\frac{e}{c}\:   \dot\vec{r} \cdot \vec{A}   + 2m \alpha  \left [
 \hat z\cdot  \left ( \dot\vec{r}  \times \vec{S}\right ) \right ] +
 \frac{ m \alpha^2}{\hbar^2} +  \omega_c S_z \: .  
 \label{sette} 
\enea
In Eq.(\ref{sette}) we have introduced the constraint that the orbital electron
motion takes place along a one-dimensional circle by parametrizing the
trajectories with the angle $\varphi$ as $ x = R \cos \varphi ; \; y =
R \sin \varphi $.
The additional {\sl Berry phase } term \cite{aurbach} in
Eq.(\ref{prospin}) arises from the fact that different spin coherent
states are not orthogonal to each other since, to leading order in
$\epsilon$, the scalar product between spin-coherent states at times
$t_j , t_j +\epsilon $, $ | \Omega ( t_j ) \rangle$ and $ | \Omega (
t_{ j } + \epsilon ) \rangle$ is given by
\beq
\langle \Omega ( t_j + \epsilon ) | \Omega ( t_j ) \rangle \approx 
\exp \left[ \frac{i}{2} [ 1 - \cos \Theta ( t_j ) ] \dot{\Phi} ( t_j ) 
\epsilon \right]
\:\:\:\: . 
\label{ringostar}
\eneq

Let us look for the trajectories in orbital and spin space which make
the action stationary.  This requires solving the Eulero-Lagrange
equations for the Lagrangian $ {\cal L} [ \varphi , \dot{\varphi} ;
\Theta , \dot{\Theta} ; \Phi , \dot{\Phi} ]$ appearing in Eq.(\ref{propo}),
which are given by
\begin{eqnarray} 
\frac{\dot{\Theta}}{2} \sin \Theta  + \gamma 
\dot{\varphi} \: \sin  \chi \:  \sin  \Theta  &=& 0 
\label{em1}\:, \\
\sin  \Theta  [- \dot{\Phi} + \omega_c ] +
2  \gamma  
\dot{\varphi} \: \cos \Theta\: \cos \chi  &=& 0 
 \label{em2}\:, \\ 
\ddot{ \varphi} - \frac{ \gamma}{2\tau  _0} [ \dot{\Theta} \: \cos \Theta
\: \cos \chi + \dot{\Phi} \: \sin \Theta\:  \sin \chi  ]
&=& 0 \label{em3}
\: ,
\end{eqnarray}

with $\chi = \varphi -\Phi $. 

In order to extract the relevant physics from the above equations, we
multiply Eq.(\ref{em3}) by $\dot{\varphi}$, and, by use of
Eqs.(\ref{em1},\ref{em2}), we rewrite Eq.(\ref{em3}) as:
\beq \dot{\varphi} \ddot{
\varphi} =- \frac{ \omega _c}{4\tau _0} \: \dot{\Theta} \: \sin\Theta
\:\:\:\: .
 \label{orbo} 
\eneq
A straightforward time integration gives:
\beq
\tau_0 \frac{ ( \dot{\varphi} )^2 }{2} - \frac{\omega_c}{4} \cos \Theta = cnst
\:\:\:\: , 
\label{orbo2}
\eneq

which states that the total energy is conserved. According to
Eq.(\ref{orbo2}), the particle energy only includes the orbital kinetic
term and the Zeeman term. This is obvious, since the force associated
to the spin orbit coupling, being gyroscopic, does no work.
A change in the precession angle $\Theta$ implies acceleration in the
orbital motion.  According to the spin Hamiltonian of Eq.(\ref{three})
the RSOI is responsible for flipping of the spin, while the Zeeman
coupling tends to stabilize the spin direction.  Hence, there are two
physically relevant limits, in which the orbital motion fully
decouples from the spin dynamics, according to the inequality
$\omega_c/2 <(>) \dot \varphi$, as we are going to show next.
Both limits involve  an orbital  motion with constant velocity 
$\dot \varphi $ and a spin orientation  given by the angles 
$\Theta = {\rm constant} \:\:\: , \:\:
\varphi - \Phi = 0 \: (mod.\: \pi ) $.

{\sl a) Vanishing Zeeman coupling}:  This case has already been considered 
 and it has been shown that it can be exactly solved
analytically\cite{tserkov}. In this case Eq.(\ref{orbo})
allows for the semiclassical solution $\dot\varphi =cnst $ and quantum
fluctuations of the spin do not interfere with the orbital motion.
 The spin is tilted by the constant angle $\tan \Theta = 2\gamma $ and 
precesses with a constant frequency $\dot \Phi = \dot \varphi $. 

{\sl b) Large Zeeman coupling}: flipping of the spin and quantum
fluctuations of the spin are strongly disfavoured, again corresponding 
to the classical
saddle point configuration $\Theta = {\rm constant} \:\:\: , \:\:
\varphi - \Phi = 0, \pi  $  with 
\beq
\pi/t_f=\dot{\varphi} = \pm \dot{\Phi} = 
 \omega_c \:\frac{1}{1-2\gamma \:  {\rm ctan} \Theta}\:\:\:.
\label{orbo3}
\eneq
The spin precesses around the $z$-axis, with the same angular velocity
as the orbital motion. When $\gamma \to 0 $, $\Theta $  can  be
vanishingly small, with $\dot\Phi =\omega _c  $.

It may appear that the simple spin precession provided by
Eq.(\ref{orbo}) with $\Theta = cnst $ is a solution of
Eq.s(\ref{em1},\ref{em2},\ref{em3}) for any ratio $\omega _c
/\dot\varphi $. Were this the case, there would be no need to invoke
the limitations of case $b)$ as stated above.  However, a careful
analysis of the stability of this saddle point solution shows that the
simple spin precession is a minimum of the action only when the Zeeman
coupling is strong. In particular, if $\omega _c /\dot\varphi >>1 $,
an additional condition $\omega _c \tau _o/\gamma ^2 >1 $ has to be
satisfied. Otherwise the frequency of the fluctuations around the
saddle point solution does not keep real and the analysis of small
oscillations in the parameter space around the saddle point solution
breaks down.

This shows that a classical orbital motion with constant velocity is
compatible with both limits of large and small ratios of the Zeeman
coupling to the RSOI strength. However the two limiting cases do not
belong to the same saddle point. In particular there will be a
crossover region connecting the two limits in which quantum
fluctuations of the spin may induce changes in the orbital velocity of
the particle.

In the rest of the paper we will choose $\dot \varphi=const$ piecewise
and parametrize the quantum dynamics of the spin with the value of the
velocity obtained by the stationary phase condition discussed in the
next section.  As discussed above, our approximation cannot reproduce
faithfully the intermediate region of parameters ranging the two
limits of large and small Zeeman coupling $\omega_c / \gamma
\dot\varphi <(>) 1$.  We checked numerically the reliability of our
approximation, by numerically integrating
Eqs.(\ref{em1},\ref{em2},\ref{em3}) and have have found that it is
satisfactory in most of the parameter range.

By putting $\dot\varphi =cnst $ into Eqs.(\ref{em1},\ref{em2}), they
become, as we show in detail in Appendix A, the classical equations of
motion for a magnetic moment in a time dependent magnetic field
${\cal{B}}
\equiv ({\cal{B}}_+,{\cal{B}}_-,{\cal{B}}_z ) = \left ( \gamma
\dot{\vt} \: e^{i\vt}, \gamma \dot{\vt} \: e^{-i\vt},-\omega _c/2
\right ) $. This can be easily understood from the fact that
minimizing the action in Eq.(\ref{propo}) directly w.r. to the spin
components provides:
\beq 
\vec{S}\times \frac{d\vec{S}}{dt} = - \frac{\delta H }{\delta
\vec{S}} \:\:\:\: , 
\eneq 
which has to  be solved together with the constraint of constant 
modulus: $\vec{S}\cdot d\vec{S} /dt = 0$ ( see Eq.(\ref{appe1}).

Among the other possible saddle-points, a solution of the motion
equations can be found with the particle trapped within the ring arm
(turning points of $\varphi(t)$ are at $\varphi = 0,\pi $) if $\gamma
$ is large enough. This solution doesn't seem to be practical as it
requires fine tuning of the external parameters with transfer of
energy from the spin motion to the orbital motion and viceversa. We
have not investigated it in detail, but we expect that it could
provide resonant tunneling across the ring.
\section{Saddle point approximation and looping in the ring}
In this Section we show how we implement the piecewise saddle point
solution for the orbital motion, $\dot{\vt} = cnst$, to study the
coherent propagation of the electron inside the ring.
In the following, we will denote by ``loop'' and ``looping
trajectory`` both trajectories that wind around the ring (closed), and
paths in which the particle moves forth and back in one of the ring
arms (open) (see Fig.(\ref{WLpaths})).  Of course, the amplitudes
differs very much in these two types of looping . Indeed, the net spin
rotation is different between the two paths and, also, the
Ahronov-Bohm ($AB$) phase is absent in the latter ones .
\begin{figure}[!htp]
    \centering \includegraphics[width=\columnwidth]{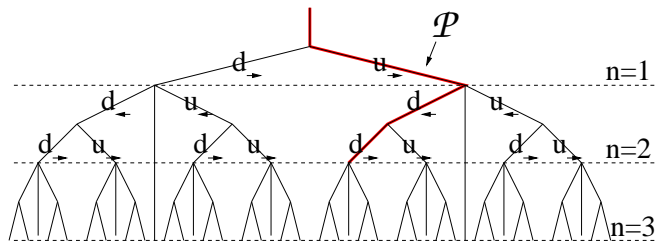}
\caption{(color online) Cayley tree describing the way in which higher
  order paths are built in the numerical code. Full lines correspond to
  propagation in the upper arm ( $u$ ) or lower arm ( $d$) in the
  forward ($\rightarrow$) or backward ( $\leftarrow$) direction. The
  nodes represent the leads of the ring. The exit nodes are marked by
  a broken line at each order $n$.  The heavy line correspond to the
  path reported in Eq.(\ref{ugly}). }
\label{tree}
\end{figure}
In general, at order $n$, we will have $2^{(2|n|-1)}$ trajectories of
a particle entering the ring at $\vt = 0 $ and exiting at $\vt = \pi $
and each of them will include $2|n|-1 (n = \pm 1,\pm 2,...) $
elementary paths (or ``stretches'') of the type
$u_\rightarrow,u_\leftarrow,d_\rightarrow,d_\leftarrow$, as classified
at the end of Section II.  Looping at order $|n|$ involves $2 |n|$
scattering processes at the contacts.  In the case of spinful
electrons the $S$-matrix is doubled ($6\times 6$) w.r. to the one
given in Eq.\ref{matr}.  Of course, the $S$-matrix at the contacts is
sample dependent and any special choice is arbitrary.  In the
following we neglect possible asymmetries in the up-down channel as
well as accidental spin flipping in traversing the contacts. Hence, we
make the simplifying assumption that the $S$-matrix is block diagonal
of the form given in Eq.\ref{matr} for both spins.

Each time the trajectory impinges at a contact without leaving the
ring, the $S-$ matrix of Eq.(\ref{matr}) alledges for either forward,
or backward scattering.  Let us denote with
$u_{\rightarrow(\leftarrow)}(t_i,t_j)$ the forward (backward)
propagation amplitude in the upper arm from time $t_j$ to time $t_{i}$
($d_{\rightarrow(\leftarrow)}(t_i,t_j)$ for the lower arm).  The
$u(d)$ amplitudes include the Ahronov-Bohm phase and the spin
evolution, but not the dynamical phase, which is factored out (see
below).  As shown in section II, to first order ($|n|=1$) there are
only two possible paths (see Fig.(\ref{noWLpaths} 1a,1b)), the
propagation amplitude is, then, the sum of the the corresponding two
amplitudes:
\bwt
\beq
A_1(\mu _f ; \mu _0 | E_0 ) = \int_0^\infty \: \frac{d t_f}{ \tau_0} 
e^{ i \frac{E_0 t_f}{\hbar}}  
\left\langle\mu_f,t_f\right|\biggl [ ( \mathcal S_{3'1} u_\rightarrow(t_f,t_0) \mathcal S_{13} +  \mathcal
S_{3'2} d_\rightarrow(t_f,t_0) \mathcal S_{23} ) \biggr ] 
\left|\mu_0,t_0\right\rangle \: 
e^{-i\frac{mR^2}{2\hbar }  \int _{t_0}^{t_f} dt \: \dot \vt ^2 (t )}
\label{prio}
\eneq
The amplitudes have to be summed all together, order by order.  The
key observation appearing in the symbolical writing of Eq.(\ref{ugly})
is that the dynamical phase, at a given order, does not depend on the
chirality of the motion and can be factored out.  On the contrary, the
Ahronov-Bohm phase depends on the chirality, and the spin evolution
depends on both the chirality and on the modulus of the propagation
velocity.
Beyond the first order, there is a net increase in the number and type
of the paths to be summed together. In Fig.(\ref{tree}), we
pictorially sketch all the possible paths by mean of a Cayley tree.
Each node represents a lead of the ring and, according to the
scattering matrix of Eq.(\ref{matr}), the electron can be
backreflected into the ring's arm it is coming from with an amplitude
$S_{ii}$, or, with an amplitude $S_{ij}$, it can be either transmitted
to the other arm, or outside of the ring.  In the tree, the
transmission out of the ring is possible at all the nodes crossing the
dashed lines. Each dashed line is labeled by the order $n$ of the
interference in the ring. As an example, we explicitely write down one
of the possible second order paths (the bold red line marked by $\cal
{P}$ in Fig.(\ref{tree}), which corresponds to
Fig.(\ref{WLpaths},2h)):
\bea
 A_2^{\cal{P}} (\mu _f ; \mu _0 | E_0 )=\!
\int_0^\infty \!\! \frac{d t_f}{ \tau_0} \! e^{ i \frac{E_0 t_f}{\hbar}}  
\int _{t_0}^{t_f} \!\!\!dt_2 \! 
\int _{t_0}^{t_2}\!\! dt_1 \!    
\left\langle\mu_f,t_f\right|  
\mathcal S_{3'2} d_\rightarrow(t_f,t_2) \mathcal S_{22} d_\leftarrow(t_2,t_1) \mathcal
S_{21} u_\rightarrow (t_1,t_0)\mathcal S _{13}
\left|\mu_0,t_0\right\rangle
e^{-i\frac{mR^2}{2\hbar }  \int _{t_0}^{t_f} dt  \dot \vt ^2 (t )} \:.
\label{ugly}
\enea

At the saddle point with uniform velocity, $\dot{\vt} = 2 \pi \:
(2|n|-1) / t_f $, we classify the collection of sequences belonging to
a certain $n$ as $\{{\cal{C}}_n\} $, and denote the superposition of
amplitudes (e.g., the ones in the big parenthesis of Eq.(\ref{prio})
to first order ) as $ {\cal{F}}
\left [\mu _f, \mu _0; q, \dot{\vt} | {\cal{C}}_n \right ] $.  In this
way, we can rewrite the full amplitude of Eq.(\ref{ampclas1}) as:
\beq
\mathcal A(\mu _f ; \mu _0 | E_0 )  = 
\int_{0}^{\infty} \: d t_f  \:  e^{-i E_{0}
t_f/\hbar}
\sqrt{\frac{\tau _0}{\pi i t_f } } \:
 \: {\sum _{n=-\infty}^{+\infty }}' \:  \sum _{\{{\cal{C}}_n\}}
 \: {\cal{F}} \left [ \mu _f, \mu _0; q,  \dot{\vt} = 2 \pi \: (2 |n|-1) / t_f
 |  {\cal{C}}_n \right ]   \:
e^{  -i  \pi ^2 (2|n| -1) ^2 \tau _0 / t_f  }
 \: .
\label{coll}
\eneq 
\ewt
 The series in Eq.(\ref{coll}) is  uniformly convergent. Thus, we may
swap the integral with the sum, and integrate term by term.  
The integral contributing to order $n$ is given by
\bea I_n = \int_{0}^{\infty} \: d t_f \:
\sqrt{\frac{\tau _0}{\pi i t_f }} \: e^{-i E_{0} t_f/\hbar -i \pi ^2
(2|n| -1) ^2 \tau _0 /t_f } \: \nonumber\\ =\: \sqrt{\frac{1}{\pi i }}
\: \int_{0}^{\infty} \: d x \: e^{-i\epsilon \: x^2 -i \pi ^2 (2|n|
-1) ^2 /x^2 }
\label{stat}
\enea 
with $\epsilon = E_{0} \tau _0/\hbar $.  We compute it approximately
within stationary phase contribution. Since the phase of the exponent
of the integrand is stationary at $\bar{t}_n = \epsilon ^{(-1/2)} \:
\pi \:(2|n|-1) \tau _0$, by inserting this value in the phase and
integrating out the gaussian fluctuations, we readily get:
\bea
I_n \approx  e^{-i \sqrt{\epsilon}  2 \pi (2|n|-1) } 
\sqrt{\frac{1}{\pi i  }} \:
  \int_{-\infty}^{\infty} \: d (\delta x ) \:
  e^{-i \epsilon \:( \delta x )^2} \nonumber\\
  \sim 
\frac{1}{i \sqrt{2\epsilon  }}
 \: e^{-i \sqrt{\epsilon}  2 \pi  (2|n|-1)^2  } \:\:\: .
\enea 
($ -i \epsilon ^{-1/2} $ is the usual factor appearing in the
one-dimensional free particle Green's function in real space and
energy).  This approximation, when plugged into Eq.(\ref{coll}),
provides the final result:
\bwt
\beq
A(\mu _f ; \mu _0 | E_0 )  = 
 \frac{1}{i \sqrt{ 2\epsilon } } \: 
 {\sum _{n=-\infty}^{+\infty }}' \: \biggl (  \sum _{\{{\cal{C}}_n\}}
 \: {\cal{F}} \biggl [ \mu _f, \bar{t}_n , \mu _0; q,  \dot{\vt} =
 2 \pi \: ( 2|n|-1 ) / \bar{t}_n
 \biggl | \biggr .   {\cal{C}}_n \biggr ]  \: \biggr )  \: 
    \: e^{-i \sqrt{\epsilon}  2 \pi  (2|n|-1)  }
 \: .
\label{colf}
\eneq 
\ewt
The enumeration of the trajectory configurations belonging to the
collection $ {\cal{C}}_n $, to order $n$, is numerically performed
order by order.

In the next Section, we discuss the elementary
propagators for each of the four stretches,
$u_{\rightarrow},u_{\leftarrow},d_{\rightarrow},d_{\leftarrow} $, as
defined in Section III.  This allows us to construct the functional $
{\cal{F}} $ for each incoming and outgoing spin polarization.

\section{Quantum spin  dynamics of the electron  propagating in the ring} 
In this Section, we provide the explicit formula for the spin
contribution to the total propagation amplitude, given by
\beq
\hat{U}_{\rm spin} ( t , t' ) = {\bf T} \exp \left[ - \frac{i}{ \hbar}
\int_{ t' }^{ t } \: d \tau \: H_{\rm spin} ( \tau ) \right] 
\:\:\:\: .
\label{a.2.1}
\eneq

As discussed in detail in appendix A, within the saddle point
approximation, $ H_{\rm spin} ( t )$ corresponds to the Hamiltonian of
a spin-1/2 in a time-dependent external magnetic field. It can be
written as (from now on, we will denote by $\omega_o$ the frequency of
the orbital motion, that is, the stationary phase value of
$\dot{\varphi}$)

\beq
\hat{H}_{\rm spin} ( t ) = \left[ \begin{array}{cc} r \cos  \vartheta  & 
r \sin  \vartheta  e^{ - i \omega_o t } \\
r \sin  \vartheta  e^{  i \omega_o t  } &
- r \cos  \vartheta  \end{array} \right]
\label{three_1}
\:\:\:\: , 
\eneq

with
\begin{eqnarray}
 r \cos \vartheta =   \frac{ \omega _c}{ 2}\: ,  \:\:\:\:\:
r \sin \vartheta = \gamma \omega_o \: ,   \:\:\:\:\:
\varphi (t) =  \omega_o  t \: ,    \label{def}\\ 
r =  \met \sqrt{ \omega _c^2+4 \gamma ^2  \omega_o ^2} \: ,
  \:\:\:\:\: \tan \vartheta  =   \frac{2\gamma  \omega_o }{ \omega _c}
\:\: .
 \nonumber
\end{eqnarray}

Including only the AB phase implies $\vartheta = 0 $, while including
only RSOI implies $\vartheta \to \pi /2 $.

It is useful to solve for the spin dynamics in the representation of
the instantaneous eigenstates of $\hat{H}_{\rm spin} (t)$. At fixed
$t$, its eigenvalues are given by $\pm \epsilon = \pm r$, while the
corresponding eigenvectors take the form:
\begin{eqnarray}
| + , t \rangle = \left ( \begin{array}{c} 
    \cmt  \\  
 \smt \: e^{\ang }     \end{array} \right ) \: ,  \:\:\:\:\:\:
| - , t \rangle = \left ( \begin{array}{c} 
    -\smt \: e^{ - \ang}  \\  
 \cmt   \end{array} \right ) \: . 
\label{four}
\end{eqnarray}

Thus, the matrix diagonalizing $\hat H_{\rm spin} ( t ) $ at time $t$
is
\begin{eqnarray}
\hat B ( t ) \equiv \left[ \begin{array}{cc} 
 \cmt &   - \smt \: e^{ - \ang} \\
 \smt \: e^{  \ang}  &  \cmt
 \end{array} \right] 
\:\:\:\: .
\label{five}
\end{eqnarray}
The matrix $\hat B ( t )$ encodes the adiabatic contribution to the
evolution of $ | \Psi ( t ) \rangle$. Therefore, in order to write
down the Schr\"odinger equation with Hamiltonian $\hat{H}$,
\[
\biggl\{ i \frac{ \partial}{ \partial t} - \hat{H} ( t ) \biggr\} 
| \Psi ( t ) \rangle = 0 
\]

in the adiabatic basis, one has to strip off from the state $ | \Psi (
t ) \rangle$ its adiabatic evolution, operating with $\hat{B}^\dagger
(t)$, so to get:
\beq
 \biggl[ i \frac{ \partial}{ \partial t} 
-   \hat{B}^\dagger ( t ) \hat{H} ( t ) \hat{B} ( t ) +  
 \hat{B} ( t )^\dagger  i \frac{ \partial}{ \partial t} \hat{B} ( t ) 
\biggr]  \hat{B}^\dagger ( t ) | \Psi ( t ) \rangle  = 0 \:\: .
\label{addi2}
\eneq

Eq.(\ref{addi2}) may be rewritten in a 2$\times$2 matrix formalism.
Let $ \left( \begin{array}{c} u_+ \\ u_- \end{array} \right )$ be the
components of $ | \Psi ( t ) \rangle$ in the adiabatic basis. The
corresponding system of differential equations reads

\beq
i \frac{d}{dt}\left ( \begin{array}{c}
    u_+  \\
    u_-     \end{array} \right ) \:=
\hat{H} _A(t) \: 
\left ( \begin{array}{c}
    u_+  \\
    u_-
\end{array} \right )\label{addi6}
\eneq
where we have defined
 
\bea
 \hat{H}_A= 
\left( \begin{array}{cc} 
r+ \wo \smqt & \met  \wo \sin\vartheta   e^{-i \wo t} \\
\met \wo \sin\vartheta  e^{i \wo t} & - r - \wo \smqt \end{array}
\right)\:.
\label{berryham}
\enea
The  extra  term  appearing on the diagonal w.r.to the hamiltonian of 
Eq.(\ref{three_1}) is just the Berry  phase:
\beq
 \langle + , t | i \frac{d}{ d t } | + , t \rangle = -
\langle - , t | i \frac{d}{d t } | - , t \rangle =  \omega _o \: \smqt \:\: . 
\eneq
Eq.(\ref{addi6}) is solved in Appendix B  and   the full propagator 
in the representation  of the instantaneous eigenvectors  reads: 
\bwt 

\beq
U(t,t')\:=\:
\left ( \begin{array}{cc}
(\cos(\epsilon (t-t'))- i \eta \sin(\epsilon (t-t')))e^{i/2 \varphi
 (t-t')}  &
-i \beta \sin (\epsilon (t-t'))e^{i/2 \varphi(t+t')}   \\
 -i \beta \sin (\epsilon (t-t'))e^{-i /2  \varphi (t+t')}    &
(\cos(\epsilon (t-t'))+ i \eta \sin(\epsilon (t-t'))) e^{-i/2  \varphi
 (t-t')}
 \end{array} \right )\;.
\label{propagatorespin}
\eneq
\ewt
where  $ \epsilon=\pm\sqrt{ (r + \frac{\wo}{2}\cos
\vartheta )^2+s^2}$  and $s = \frac{\omega _o}{2} \sin \vartheta $. Also:
\[\beta=\frac{\wo}{2\epsilon } \sin\vartheta \:\:\: , 
\:\:\:\eta =\frac{ r + \frac{\wo}{2}\cos
  \vartheta  }{ \epsilon}\: . \]
This is the propagator in the adiabatic basis.  Therefore, in order to
switch to the fixed  spin basis, one should write
$U_{spin}(t,t')=B(t)U(t,t')B^\dagger(t') $, where $B(t) $ is given by
Eq.(\ref{five}).
The four elementary  stretches imply  the following substitutions in 
 the propagator  of Eq.(\ref{propagatorespin}):
  
 $u_\rightarrow$)  {\sl   forward orbiting in the upper arm of
the ring :}   $\:\:\:\:\:\:\varphi  (t) =   \omega _o t $ and  $\vartheta \to  \vartheta $.

 $ u_\leftarrow$)  {\sl   backward orbiting in
the upper arm of the ring :}    $\:\:\:\:\:\:\varphi  (t) =  \pi  - \omega _o t $ and 
 $\vartheta \to  -\vartheta  $.

 $d_\rightarrow$)  {\sl   forward orbiting in the lower arm of the ring :} 
     $\:\:\:\:\:\:\varphi  (t) =  2\pi  - \omega _o t $ and  $\vartheta \to  -\vartheta  $.

 $ d_\leftarrow$)  {\sl  backward orbiting in the lower arm of the ring :}
    $\:\:\:\:\:\:\varphi  (t) =  \pi  + \omega _o t $ and  $\vartheta \to  \vartheta  $.
 
\section{the  conductance}
In this section, we derive the DC conductance  ${\cal{G}}$ across the ring, at
the Fermi energy.  Within Landauer's approach, ${\cal{G}}$ is given by
\beq
{\cal{G}}=\frac{e^2}{\h}\sum_{\sigma,\sigma'}
 \left| \mathcal A (\sigma;\sigma'|E_0)\right|^2
\:\:\:\: . 
\label{ec.1}
\eneq
We will here consider the dependence on the external magnetic field
($\phi /\phi _0$) and on the spin-orbit strength $k_{SO} R
$\cite{nota} both in absence and in presence of dephasing at the
contacts.  The various amplitudes in Eq.(\ref{ec.1}) have been
numerically computed, as discussed in Sec.s(III-VI).
\begin{figure}[!htp]
    \centering \includegraphics[width=\columnwidth]{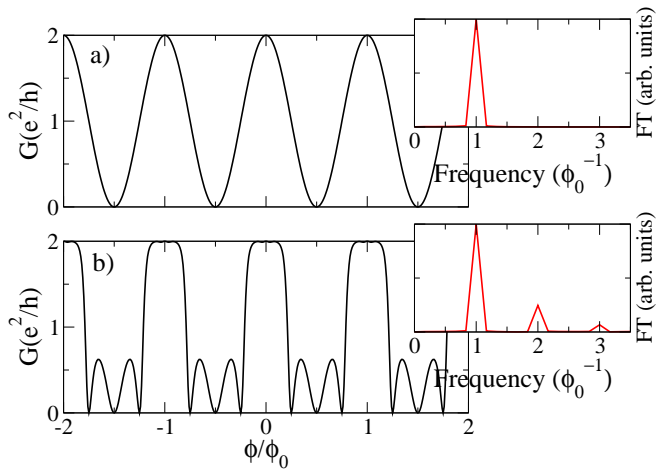}
\caption{(color online) $a) $ Magnetoconductance of an ideal ring as a
  function of the magnetic field $\phi/\phi_0$. The Fourier transform
  (FT) of the magnetoconductance ({\sl in the inset}) shows only the
  AB peak at freq. $\phi_0^{-1} $.  $b) $ Magnetoconductance of a
  realistic ring in which the S matrix of Eq.(\ref{matr}) regulates the
  scattering of the electron at the leads. As a consequence of the
  backscattering, the FT of the magnetoconductance ({\sl inset})
  shows higher order frequencies.}
\label{GvsB}
\end{figure}
In Fig.(\ref{GvsB}) we show the magnetoconductance of the ring in the
absence of RSOI ($k_{SO}R =0$).
In panel a) of Fig.(\ref{GvsB}), only the path of
Fig. (\ref{noWLpaths}, $1a)$ has been  considered,
 i.e., full transmission across the ring is
assumed, as it would be the case for ideal coupling to the leads.  The
corresponding Fourier spectrum is showed in the corresponding inset. 
The well known
Ahronov-Bohm sinusoidal pattern implies that just the fundamental
frequency $\phi_0^{-1}$ appears.

To make the model more realistic, we allow for higher order looping of
the electron within the ring.  In Ref.\cite{noiletter}, only the paths
of the kind of Fig.(\ref{noWLpaths},2a),\ref{noWLpaths},2b)) were
included.  Here, we consider also the paths of the kind of
Fig.(\ref{WLpaths}) in which the electron can be backscattered at the
leads.  We  use here $\bar r=0$ in the scattering matrix between the
arms and the leads, which means that no backreflection in the incoming
lead is present. The magnetoconductance of the ring, in this regime,
is showed in Fig.  (\ref{GvsB}b). Because of the inclusion of time
reversed  paths (TRP) within the ring, we see that higher
order frequencies appear in the Fourier spectrum. In particular, the
inset shows a peak at $2/\phi_0$ which is the signature of weak
localization\cite{nonloso}.
\begin{figure}[!htp]
    \centering \includegraphics[width=\columnwidth]{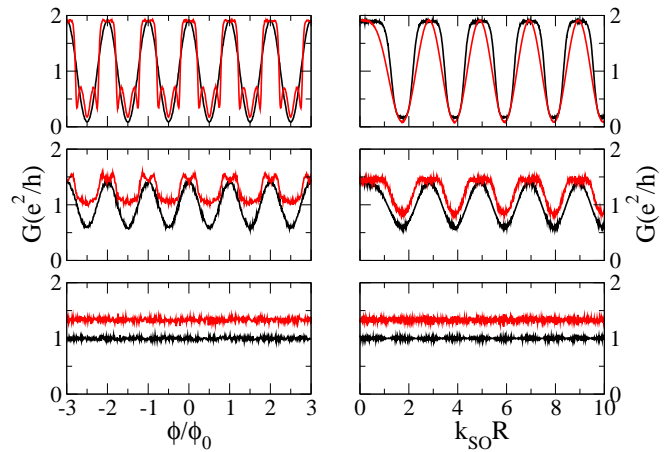}
\caption{(color online)Conductance as a function of $\phi/\phi_0$ {\sl
(left panels)} and $k_{SO}R$ {\sl (right panels)} for ideal {\sl
(black curves)} and realistic {\sl (red curves)} contacts. An
increasing amount of dephasing at the contacts is also included: {\sl
from top to bottom:} $\zeta=\pi/3,\pi,2\pi$}.
\label{GvsBSO}
\end{figure}
We also include some dephasing due to diffusiveness in the contacts by
adding a random phase $z\in (-\zeta,\zeta ) $ for each scattering at
the leads.  In Fig.(\ref{GvsBSO}), we report the conductance {\cal{G}}
{\it vs.}  $ \phi / \phi_0$ , with $k_{\text {SO}}R = 0$ {\sl (left
panels)} or {\it vs.}  $k_{SO} R$ with $\phi/\phi_0=0$ {\sl (right
panels)}. These are averaged over $N=1000$ realizations of disorder,
and plotted increasing the window of phase randomness ($\zeta = \pi/3
,\pi, 2\pi$ from top to bottom). The black curves refer to the case of
Fig(\ref{GvsB} $a)$) (ideal contacts) while the red curves refer to
the case of realistic contacts (Fig(\ref{GvsB} $b)$), with
$\bar{r}=0$.

By comparing the top left panel of Fig.(\ref{GvsBSO}) with
Fig.(\ref{GvsB}), we see that the ring is rather insensitive to small
dephasing at the contacts. By increasing the amount of dephasing 
{\sl (middle and bottom left panels in Fig.(\ref{GvsBSO}))}  we find that the
 sensitiveness is larger  in the case of realistic contacts.
 This is due  to the fact that for realistic coupling, the
electrons in the ring can experience higher order paths, since it scatters
 with the leads many times.  

In the right panel of Fig.(\ref{GvsBSO}), we plot the DC conductance
{\it vs.}  $k_{\text {SO}}R$ at $\phi/\phi_0 =0 $ for both ideal
contacts and realistic contacts (and $\bar{r}=0$) {\sl (black
and red lines in each box)}, with an increasing phase randomization 
{\sl (boxes
from top to bottom with $\zeta = \pi/3 ,\pi, 2\pi$)}, averaged over
$N=1000$ realizations. In the case of ideal contacts and little
dephasing {\sl (top right panel black curve)}, we see again the
quasiperiodic oscillation of the conductance reproducing the
localization conditions at the expected values of $k_{\text {SO}}R$
\cite{frustaglia,molnar,souma,dario,noiletter}.  When including higher
order processes, interference effects give rise to a slightly
different pattern.  In the case of realistic contacts, we note that
 the device is seriously  affected by dephasing, mainly because
including  the TRPs contributing to the transmission   amplitude
 increases the number of scattering processes
at the leads. Indeed, when  the 
dephasing is  quite large, it gives rise to
 random oscillations that  are not
averaged out,  so  that they  wash  out the conductance oscillations.
 The effect takes 
place  for  
$\zeta\sim\pi$  when TRPs are included, in contrast to $\zeta \sim 2\pi$ 
when  the TRPs are  absent.  As regular magnetoconductance oscillations 
are experimenally
observed \cite{nitta,yau,morpurgo,kato,molenkamp} with  little
precentage of contrast between maxima and minima, we conclude that, in
real samples, dephasing is ubiquitous.

\section{Spin Transmission}

\begin{figure}[!htp]
    \centering
    \includegraphics[width=\columnwidth]{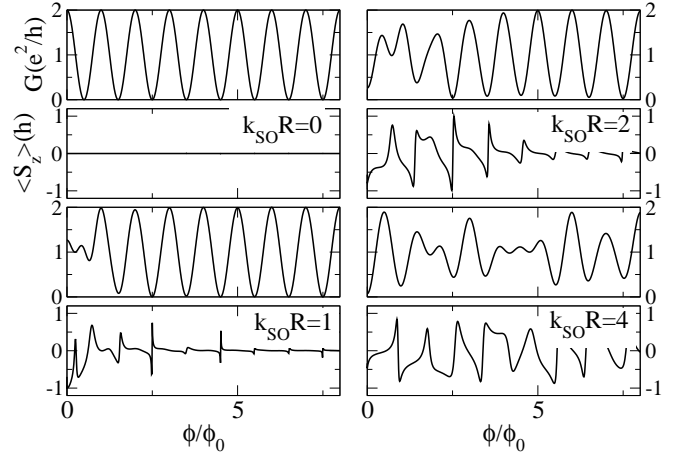}
\caption{Magnetoconductance and expectation value of the outgoing $\hat
z $ spin component for an incoming spin in the $x$ direction at
different values of the RSOI strength ($k_{SO}R = 0,1,2,4 $ indicated
in the pictures).}
\label{spintramiscela}
\end{figure}

\begin{figure}[!htp]
    \centering \includegraphics[width=\columnwidth]{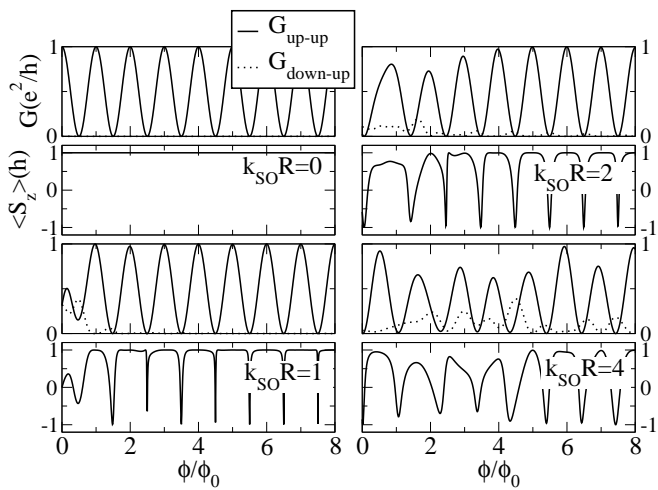}
\caption{Separate contributions to the magnetoconductance for
different outcoming spin polarizations, $ G_{up-up} $ ({\sl full line})
and $ G_{down-up} $ ({\sl dotted line} ) compared to the expected
value of the outgoing $\hat z $ spin component. The incoming spin is
polarized ``up'' .  Different values of the RSOI strength are reported
($k_{SO}R = 0,1,2,4 $ indicated in the picture ).}
\label{spintraup}
\end{figure}

In this Section we calculate the rotation of the spin of the electron
transmitted through the ring.  We first consider an incoming electron
beam with in-plane spin polarization (let's say, polarized along the
$x$ direction ). The spin rotation is measured by calculating 
 the average value of the outgoing spin:
\beq
\langle S_z\rangle= \frac{\left\langle\Psi_{out}\right|S_z\left|\Psi_{out}\right\rangle}{\left\langle
\Psi_{out}\right.\left|\Psi_{out}\right\rangle }\;.
\label{szout}
\eneq
Since in the previous Section we have shown that higher order looping
just adds subleading higher order harmonics to the conductance, here we
focus on the case of ideal contacts, that is, we include in the
calculation only paths as the ones of Fig.(\ref{noWLpaths},1a).  The
in plane polarization can be considered as a superposition of equal
weighted z-polarized spin components.  In the absence of RSOI
($ k_{SO} R =0 $),
opposite spin polarizations do not interfere with each other. As a
consequence, the total expected $\langle S_z \rangle $ component keeps
zero at the exit.  Fig.(\ref{spintramiscela}) shows the
magnetoconductance for increasing  $ k_{SO} R$, and the
corresponding expected spin component polarized along the $z-$axis 
 at the exit of the ring. The
Zeeman term is on the diagonal of the spin  Hamiltonian of Eq.(\ref{three_1})
 and operates to keep the $z-$components of the spin polarization fixed,
 while the RSOI is offdiagonal and tends to favor inplane spin polarization.
This implies that when the magnetic field increases 
($\phi /\phi _0 >> k_{SO} R$)  the spin polarization gets frozen  to the
 incoming polarization.
This is the reason why,in Fig.(\ref{spintramiscela}), at high fields,
  the   trasmitted spin polarization is inplane. Incidentally  we observe 
that this result should not be expected  in real systems in which spin 
relaxation can occur due to electron-phonon  interaction or other mechanisms
 as hyperfine interaction with nuclear spins.
Spin relaxation would induce flipping of that spin component that is 
energetically  unfavourable and the final
transmission of the spin will result  to be partly out of the $x-y$ plane.
 On the contrary,
when  $\phi /\phi _0 \sim
k_{SO} R $ the competition of the Zeeman and the RSOI induces  
a coherent rotation of the  spin while the electron travels
along the ring. 
Fig.(\ref{spintramiscela})  shows that, when $\phi /\phi _0 \sim
k_{SO} R$, the spin is moved significantly out of the $x-y $ plane,
consistently affecting the AB oscillations.
To better understand what happens when $\phi /\phi _0 \sim k_{SO} R$,
we isolate the  spin ``up'' polarization for the incoming particle 
in what follows ($\left\langle\Psi_{in}\right|S_z\left|\Psi_{in}\right\rangle/\left\langle
\Psi_{in}\right.\left|\Psi_{in}\right\rangle=1$)  and   
we separately plot in Fig.(\ref{spintraup}) the two contributions to
the conductance $G_{up-up} $ ({\sl full line}) and $ G_{down-up} $
({\sl dotted line}), for opposite polarizations of the outgoing
electron. $G_{up-up} $ is the contribution to the conductance 
arising from the particle flux that mantains the same polarization at the exit as the incoming one, while  $ G_{down-up} $ refers to a particle flux 
having  the opposite polarizations at the exit with respect to the one 
at the entrance.  Of course, when $ k_{SO} R = 0 $, the electron spin is in
the ``up''direction for any $\phi/\phi_0$. When both RSOI and magnetic
field are present, with $\phi /\phi _0 >> k_{SO} R$, the spin
polarization is steadly in the $z-$ direction, except for sharp
reversals at flux values $\phi _0 m/2 $ ($m$ integer).  However,
$G_{down-up} $ is vanishingly small at these places, together with
$G_{up-up} $. Therefore the conductance vanishes at these points
anyhow, and the transmitted spin polarization is fully up, except for these
points.  On the contrary, in the parameter intervals characterized by
$\phi /\phi _0 \sim k_{SO} R$, both $ G_{up-up} $ and $ G_{down-up} $
are non vanishing (see Fig.\ref{spintraup}), so that the spin is rotated 
at the exit, with
nonvanishing transmission amplitude.

\begin{figure}[!htp]
    \centering
    \includegraphics[width=\columnwidth]{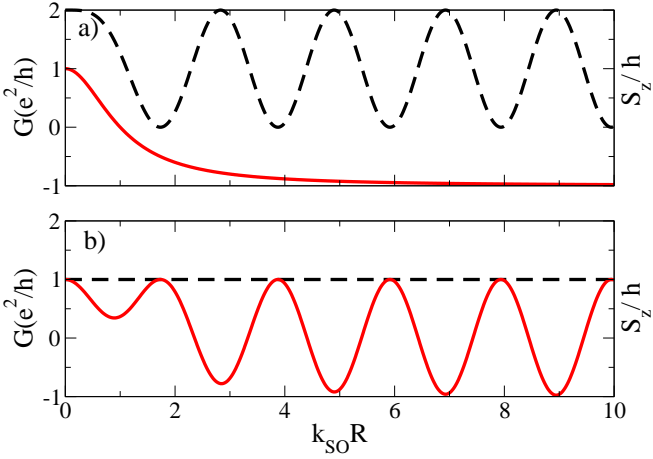}
\caption{ (color online) Conductance ({\sl broken line}) and spin
polarization ({\sl full line}) of the outgoing electron for $B=0$, as
a function of the RSOI. The spin of the incoming electrons is
polarized ``$up$'' : $a)$ The ring case with ideal contacts. $b)$ a
single wire of the same length and curvature as one of the ring arms.}
\label{spinwire}
\end{figure}

We now examine in more detail the expected dependence of the outgoing
spin polarization on the RSOI, at zero magnetic field.  For an
incoming electron polarized with spin ``up'', Fig.(\ref{spinwire}(a))
shows that a large enough RSOI produces a rotation of the spin which
points down at the exit for any $k_{SO}R$ value. Meanwhile, the total
conductance oscillates with $k_{SO}R$. This finding also appears in
Ref.\cite{souma} and is quite remarkable, because it is the
consequence of the interference between the two arms of the ring.  In
order to point out the role of quantum interference, we discuss here,
for comparison, what happens by transporting a spinful electron along
a single arm of the ring (the upper one).  When just one arm is
considered, (see Fig(\ref{spinwire}b)) spin polarization oscillates,
as a function of $k_{SO} R $ \cite{datta}, while the conductance is
always unitary, because of the conservation of the particle flux. In
the limit of large $k_{SO} R$ the propagation amplitude for the
travelling electron acquires a simple analytical form. From
Eq.(\ref{propagatorespin}) we see that, in the representation of the
instantaneous spin eigenvectors, the spin propagator at the exit time
$t_f$ (with $\omega _o t_f = \pi )$, in this limit is diagonal:
\beq
U^{u\rightarrow}(t_f,0)=
\left(
\begin{array}{c c}
i e^{-i\pi\gamma}    &  0\\
0 & -i e^{i\pi\gamma}   
\end{array}
\right) \:\: ,
\eneq
so that the spin appears not to be rotated at the exit in the rotating
reference frame. However, in order to move from the representation of
the instantaneous spin eigenvectors to the reference basis, one has to
perform the transformation with the unitary $B$ matrix of
Eq.(\ref{five}) , with $\vartheta = \pi /2$. The spin part of the
propagator for an electron travelling into the upper arm (upper path
of Fig.(\ref{noWLpaths},1a)) is then:
\bea
U^{u\rightarrow}_{spin}(t_f,0)=
B(t_f)U^{u\rightarrow}(t_f,0)B^\dagger(0)
\nonumber\\
=\left(
\begin{array}{c c}
i \cos(\pi \gamma)    &   \sin(\pi \gamma)\\
 - \sin(\pi \gamma) & -i \cos(\pi \gamma)   
\end{array}
\right) \:\: .
\enea
If we inject $up$ electrons in the upper arm only, the expectation
value of the the outcoming $S_z$ defined in Eq.(\ref{szout}) is:
$\langle S_z\rangle=\cos(2\pi\gamma)=\cos(\pi k_{SO} R)$ (note that
for large enough $\gamma$ this result well approximates the red-full
line in Fig.(\ref{spinwire}b)). The conductance is unitary, $G=2e^2/h$
(the black-dashed line in Fig.(\ref{spinwire}b), as no interference
takes place.

We now go back to the transmission along both arms simultaneously and
examine the resulting interference.  According to the rules given
after Eq.(\ref{propagatorespin}), in the same limiting case as above,
the propagator accounting for transmission of incoming $up$ spins
across the ring is:
\beq
U^{u\rightarrow+d\rightarrow}_{spin}(t_f,0)=
\left(
\begin{array}{c c}
0    &  2 \sin(\pi \gamma)\\
 - 2\sin(\pi \gamma) & 0
\end{array}
\right) \:\: ,
\eneq
so that the spin at the exit is reversed.  In fact, in the expectation
value of Eq.(\ref{szout}), the oscillations in the numerator
compensate those in the denominator, eventually giving $\langle
S_z\rangle=-1$ (for large enough $\gamma$ this result well approximate
the red-full line in Fig.(\ref{spinwire}a)).  The conductance however
oscillates according to $G/(2e^2/h)= 2\sin^2(\pi k_{SO}R/2)$, as
plotted in the black-dashed line in (Fig.\ref{spinwire}a)).

\begin{figure}[!htp]
    \centering
    \includegraphics[width=\columnwidth]{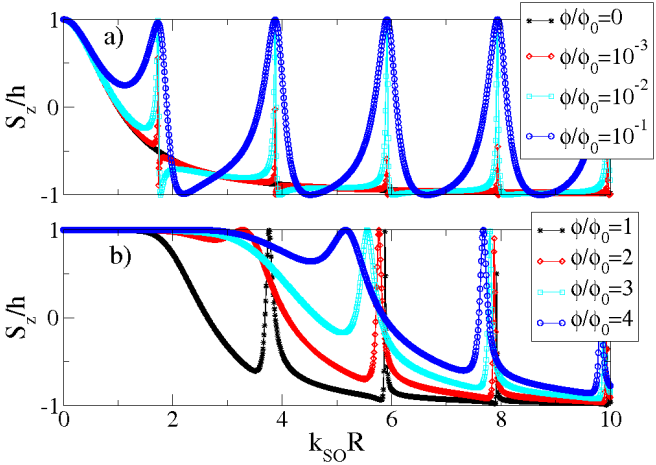}
\caption{(color online) Expectation value of the outgoing spin for
incoming spin ``$up$'' polarized electrons as a function of the RSOI
for different values of the magnetic flux : $a)$ $\phi/\phi_0$ zero or
very small; $b)$ increasing integer values of $\phi/\phi_0$. }
\label{brokensymmetry}
\end{figure}

It is quite remarkable that this result is only found at zero magnetic
field.  Indeed, no matter how small $B$ is, the time reversal symmetry
is broken and the spin oscillates with $k_{SO}R $ (see
Fig.(\ref{brokensymmetry},a).  However, for very small magnetic field
these oscillations are confined close to special values of $ k_{SO} R
= 2l $($l$ integer) and display a Lorentzian shape around these
points.  The role of the magnetic field is to make the oscillations
broader.

To summarize, there are two limiting conditions in the outgoing spin
polarization, for incoming ``up'' spin polarization: $a)$ {\sl the
zero RSOI } which leaves the incoming spin polarization unchanged,
provided no relaxation takes place; $b)$ {\sl the zero magnetic flux
case } in which the RSOI produces a flip of the spin at the exit. It
is interesting that when the flux $\phi $ is an integer number of flux
quanta $ \phi _0$, the crossover between case $a)$ and case $b) $,
with $k_{SO} R$ increasing from the value zero to values $ k_{SO} R >>
\phi /\phi _0 $ is rather sharp.  This is shown in
Fig. (\ref{brokensymmetry},b) where the expectation value of the
outcoming spin is plotted vs. $k_{SO} R$ for different integer values
of $\phi/\phi_0$.  For values of $\phi/\phi_0 >> k_{SO} R $ the
outgoing spin polarization is the same as that at the entrance, ( $up$
in the picture).  On the contrary, by increasing $k_{SO} R$ at non
zero $B$ field, we see again a pattern similar to the one of
Fig. (\ref{brokensymmetry},a), but shifted to higher values of $k_{SO}
R$.


\section{Conclusions}

To conclude, we have employed a path integral real time approach to
compute the DC conductance of a ballistic one dimensional mesoscopic
ring in both external electrical and magnetic fields orthogonal to the
ring plane.The spinful electron experiences a Rashba spin orbit
interaction and the Zeeman term. We employ a piecewise saddle point
approximation for the orbital motion, but we implement the full
scattering matrix at the leads and sum over all the possible higher
order paths up to convergency of the result. Our approach goes beyond
other recent semiclassical calculations.  Our theory is
nonperturbative and separates the adiabatic spin dynamics from the non
adiabatic one by using the rotating frame for the spin travelling
around the ring. In practice, we diagonalize the time dependent spin
Hamiltonian in the representation of the spin eigenvectors of the
instantaneous Hamiltonian.  This allows us to explore a wide range of
Hamiltonian parameters, ranging from the limit of strong magnetic
field and weak Rashba SOI to the opposite case. In both extreme
regimes our piecewise saddle point approximation is very efficient as
quantum fluctuations with flipping of the spin has little influence on
the orbital motion. This is also seen from the number of paths
required to gain full convergency.  As explained in Sec. $V$ the
separation of adiabatic from non adiabatic spin dynamics shows that in
the intermediate regimes our approximation is less justified, but
nevertheless, the results it produces seem to be in rather good
agreement with recent numerical calculations
\cite{frustaglia, molnar, shen,
dario,souma} and experiments\cite{nitta, yau,
morpurgo,kato,molenkamp}.  When we include also time reversed paths,
the Fourier transform of the magnetoconductance shows the typical
$\phi_0/2$ peak due to weak localization\cite{nonloso}.This would be the 
only surviving contribution if an ensemble average or different rings 
were measured\cite{koga}.

We have also allowed for nonideal couplings between ring and leads as
we account for dephasing effects due to diffusiveness at the
contacts. The results satisfactorily compare with experiments where
the contrast between maxima and minima in the interference fringes is
always few tens of percentage of the background DC signal.

\begin{acknowledgments}
We acknowledge financial support by the  Italian Ministry of Education
(PRIN) and by the CNR within ESF Eurocores Programme FoNE (contract
N. ERAS-CT-2003-980409).
\end{acknowledgments}

\appendix 
\section{ Motion of  a classical spin in  a rotating magnetic field}

In this appendix we derive the classical equations of motion for the
spin from the Lagrangian in Eq.(\ref{venti}), by assuming for the
orbital coordinate the saddle point solution $\dot{\varphi}$ =
constant. Once the orbital motion is dealt with in this way, the
Lagrangian for the spin degrees of freedom is given by (besides a
constant contribution)
\beq
\tilde{\cal L} [ \Theta , \Phi , \dot{\Phi } ]  / \hbar = \left( \frac{1 -  
\cos \Theta }{2} \right) \dot{\Phi} + \vec{{\cal B}} \cdot \vec{S}
\:\:\:\: , 
\label{al1}
\eneq

where  the effective time  dependent  magnetic field is 
${\cal{B}} \equiv  ({\cal{B}}_+,{\cal{B}}_-,{\cal{B}}_z ) = 
\left ( \gamma \dot{\vt} \: e^{i\vt}, \gamma \dot{\vt} \: 
e^{-i\vt},-\omega _c/2 \right ) $.   
To derive the equations of motion from a variational principle, we 
write the Berry phase term in the total spin action as
\beq
\tilde{S}_B = 
\int_{ \Phi ( 0 ) }^{ \Phi ( t_f )}  \: d t \:  \left( \frac{1 -  
\cos \Theta }{2} \right) d \Phi = \int_\Sigma \sin \Theta \: d \Theta \wedge
d \Phi 
\:\:\:\: , 
\label{al3}
\eneq

where $\Sigma$ is the spherical triangle with vertices given by the
north pole on the sphere and by the points with coordinates $( \Theta
( 0 ) , \Phi ( 0 ) ) $ , $ ( \Theta ( t_f ) , \Phi ( t_f ) )$.  Let $
( t , u ) $ be a parametrization of the spherical triangle, such that
$ \vec{S} ( t , 1 ) = \vec{S} ( t )$, and $\vec{S} ( t , 0 ) = (0 , 0
, 1)$. Thus, one may rewrite the action $\tilde{S}_B$ in
Eq.(\ref{al3}) as
\beq
\tilde{S}_{B} = \int_0^T \: d t \: \int_0^1 \: d u \: {\bf S} \cdot \left[ 
\frac{ \partial {\bf S}}{ \partial t} \times \frac{ \partial {\bf S}}{
\partial u} \right]
\:\:\:\: . 
\label{e.9}
\eneq

To derive the equations of motion, we consider a variation
$ \vec{S} ( t , u ) \rightarrow \vec{S} ( t , u ) + \delta \vec{S} 
( t , u )$ such that $\vec{S} ( T , u ) $ 
and $\vec{S} ( 0 , u )$ are "locked", that is, $ \delta \vec{S}  ( 0 , u ) 
= \delta \vec{S} ( T , u ) = 0$. 
Since $ [ \vec{S} ( t , u ) ]^2 = 1$ $\forall t , u $, one gets $\vec{
S} \cdot \frac{ \partial \vec{ S}}{ \partial t} = \vec{ S} \cdot
\frac{ \partial \vec{ S}}{ \partial u} = 0 $, As a consequence,
$\frac{ \partial \vec{ S}}{ \partial t} \times \frac{ \partial \vec{
S}}{ \partial u} $ is parallel to $\vec{S}$.  As $\delta \vec{S} \cdot
\vec{S} = 0$, this implies that
\beq
\int_0^T \: d t \: \int_0^1 \: d u \: \delta \vec{S} \cdot \left[ 
\frac{ \partial \vec{ S}}{ \partial t} \times \frac{ \partial \vec{ S}}{
\partial u} \right] = 0 
\:\:\:\: . 
\label{e.9.b}
\eneq

Thus, by integrating by parts we see that the only nonzero variation of 
$\tilde{S}_B$ is given by the boundary term 
\beq
\delta \tilde{S}_B = \int_0^{t_f } \: d t \: \delta \vec{S} ( t ) 
\cdot \left[ \vec{S} ( t  ) \times \frac{ \partial  \vec{S} ( t ) }{
\partial t} \right]
\:\:\:\: ,
\label{e.12b}
\eneq

where we have used the fact that  $\vec{S} ( t , 1 ) = \vec{S} ( t )$. 
By equating to zero the total variation of the action,  one obtains
\beq
\vec{S}\times \frac{d\vec{S}}{dt} =\vec{\mathcal B}
\:\:\:\: , 
\label{appe1}
\eneq
that is, the classical equations of motion we used in section IV.
To show that Eqs.(\ref{appe1}), when the spin components are expressed
in polar coordinates, are equivalent to Eqs.(\ref{em1},\ref{em2}), 
let us set  $\omega =  \gamma \dot{\vt} $ and $ \Omega = -\omega _c/2$. 
Also, we define
\bea
S_{+}=S_{x}+iS_{y}\hspace*{2cm}{\mathcal B}_{+}={\mathcal B}_{x} +
i{\mathcal B}_{y}\nonumber\\ S_{-}=S_{x}-iS_{y}\hspace*{2cm}{\mathcal
B}_{-}={\mathcal B}_{x}-i{\mathcal B}_{y}\vspace*{1cm}\nonumber
\:\:\:\: .
\enea
In terms of the new variables, the equations of motion are given by 
\bea
\frac{d S_{z}}{dt}=\frac{i}{2}({\mathcal B}_{+}S_{-}-{\mathcal B}_{-}S_{+})
\nonumber\\
\frac{d S_{+}}{dt}= i({\mathcal B}_{z} S_{+}-{\mathcal B}_{+}S_{z})\nonumber\\
\frac{d S_{-}}{dt}=-i({\mathcal B}_{z} S_{-}-{\mathcal B}_{-}S_{z}) \label{e2}
\:\:\:\: . 
\label{appe2}
\enea
or:
\bea
\frac{d m(t)}{dt} =i\:\left ( \Omega -\dot\varphi \right )\:p(t)-2\:i
  \omega  \:S_{z} \nonumber \\
\frac{d p(t)}{dt} =i\:\left (\Omega -\dot{\varphi} \right )\:m(t)
 \nonumber\\
\frac{dS_{z}}{dt} =-\frac{i\omega }{2}\:{m(t)}\:
\label{e4}
\enea
where
\bea
p(t)=S_{+}e^{(-i\varphi)}+S_{-}e^{(i\varphi)}\nonumber \\
m(t)=S_{+}e^{(-i\varphi)}-S_{-}e^{(i\varphi)}\nonumber
\:\:\:\: , 
\enea
and  $ 1= 4|S|^2=4 {S_{z}}^2+ (p^2-m^2 ) $.
By introducing $b= \Omega -\dot \vt $,  we obtain:
\bea
\frac{d(m(t) + p(t))}{dt}  &=&  i b \left( m(t) + p(t) \right) -2 i\omega
  \: S_z (t)   \label{m3.1}\\
\frac{d {S}_z(t)}{dt}   &=& - i \omega \:  m(t)          \label{m3.2}
\enea
Resorting to the polar coordinates  $(\Theta  ,\Phi )$ for the spin $\vec{S}$,
we get:
\bea
\left[\dTh \ct +i \left( \dP- \Omega  \right)\st \right] e^{i \chi}
 + i \ct \alpha \df=0 \label{una} \\
\left[\dTh - \omega \:  \sin{\chi}\right]  \st = 0 
\:\:\:\: . 
\label{due}
\enea
Eq.(\ref{due}) is the same as Eq.(\ref{em1}). The real part 
 of Eq.(\ref{una})  is proportional  to the imaginary part:  both give  
Eq.(\ref{em2}) when  equated to  zero,  which completes the proof. 
\section{The  spin propagator} 
In order to find the propagator of the Berry Hamiltonian $ \hat{H}_A $
of Eq.(\ref{berryham}),
 we  solve
the system of differential  Eq.(\ref{addi6}), in the 
representation of the instantaneous eigenvectors:
\bwt
\beq
i \frac{d}{dt}\left ( \begin{array}{c}
    u_+  \\
    u_-     \end{array} \right ) \:=
\left ( \begin{array}{cc}
   r+ \wo \smqt   & \wo \smt \cmt e^{-i \wo t}  \\
   \wo \smt \cmt  e^{i \wo t}&  -r- \wo \smqt
\end{array} \right )
\left ( \begin{array}{c}
    u_+  \\
    u_-
\end{array} \right )\label{addi6_1}
\:\:\:\: . 
\eneq
\ewt 
To solve Eq.(\ref{addi6_1}), first of all, we switch to a
time-independent coefficient matrix by defining:
\beq
\left ( \begin{array}{c}
    y_+  \\
    y_-     \end{array} \right ) \:=
\left ( \begin{array}{cc}
     e^{ +i \frac{  \wo }{2} t  } & 0  \\
     0 &   e^{- i \frac{\wo}{2}t   }
\end{array} \right )
\left ( \begin{array}{c}
    u_+  \\
    u_-
\end{array} \right )\:.
\label{T}
\eneq
By setting
\[Y=
\left ( \begin{array}{c}
    y_+  \\
    y_-     \end{array} \right ) \:\:\: ; \:\: 
W=\left ( \begin{array}{c}
    u_+  \\
    u_-
\end{array} \right )\:,\]
we define the matrix $T$ through
\beq Y\:=\:T\:W,\;\;\;W\:=\:T^{-1}\:Y\:.\eneq

Eqs.(\ref{addi6_1}) now read:
\bea
i \frac{ d y_+}{ d t } ( t ) = (r-\frac{\wo}{2}cos(\vartheta)) y_
+ ( t ) + \frac{  \wo}{ 2 }
\sin  \vartheta   y_- ( t ) \:,\nonumber \\
i \frac{ d y_-}{ d t } ( t ) =
+ \frac{  \wo}{ 2 }  \sin  \vartheta   y_+ ( t )
+(-r+\frac{\wo}{2}cos(\vartheta)) y_- ( t )\:.
\enea
Now we define $ r' = r - \frac{ \wo}{2} \cos \vartheta $ and $s=
\frac{ \wo}{2} \sin \vartheta $, so that in matrix form we have:
\begin{eqnarray}
i \frac{d}{dt}
\left ( \begin{array}{c}
 y_+  \\
 y_-     \end{array} \right ) \:=
\left ( \begin{array}{cc}
r' & s \\
s  &  -r'
 \end{array} \right )
\left ( \begin{array}{c}
 y_+  \\
 y_-     \end{array} \right ) \:,
\label{dicia}
\end{eqnarray}
in a compact form we can  rewrite the last equation as:
\beq
i \frac{d}{dt}Y\:=\:C\:Y\:,
\label{matr1}
\eneq
which defines the  matrix $C$.\\ We now decouple the
previous system of equation by diagonalizing the matrix $C$.
Its eigenvalues are $\lambda=\pm \epsilon=\pm\sqrt{r'^2+s^2}$ and the
matrix that diagonalizes $C$ is
\beq
P=\left ( \begin{array}{cc}
1 & \frac{r'-\epsilon}{s} \\
\frac{\epsilon-r'}{s}  &  1
 \end{array} \right ) \:.
\eneq
Its inverse is
\beq
P^{-1}=\left ( \begin{array}{cc}
 \frac{s^2}{2\epsilon(\epsilon -r')}  & \frac{(\epsilon-r')s}{2
\epsilon(\epsilon -r')} \\
-\frac{(\epsilon-r')s}{2\epsilon(\epsilon -r')}  & \frac{s^2}{2\epsilon(
\epsilon -r')}
 \end{array} \right ) \:.
\eneq

Eq.(\ref{matr1}) now reads:
\beq
i \frac{d}{dt} P^{-1}\: Y\:=\:P^{-1}\:C\:P\:P^{-1}\:Y\:,
\eneq
which, by defining $V\:=\:P^{-1}\:Y$, becomes
\beq
i \frac{d}{dt} V\:=\left ( \begin{array}{cc}
\epsilon  & 0          \\
    0     &  -\epsilon
 \end{array} \right ) \:      V\:.
\eneq
Its formal solution is:
\beq
 V(t)\:=\left ( \begin{array}{cc}
e^{-i\epsilon (t-t')}  & 0          \\
    0     &  e^{i\epsilon (t-t')}
 \end{array} \right ) \:      V(t')\:,
\eneq
or, in matrix form
\beq
V(t)\:=\:S(t-t')\:V(t')\:.
\eneq
Now we apply inverse transformations, in order to obtain the full
Schr\"odinger propagator, that is the matrix transformation between
($W(t')$ and $W(t)$).
\[
W(t)\:=\:T^{-1}(t)\:P\:S\:P^{-1}\:T(t')\:W(t')
\;\;,\]
where $P$ is time independent. 
The full evolution operator in the adiabatic basis is:
\beq
U(t,t')\:=\:T^{-1}(t)\:P\:S\:P^{-1}T(t')\:;
\eneq
By performing all the matrix products, we obtain  the result 
of  Eq.(\ref{propagatorespin}), as 
given  in  the text.

\end{document}